\documentclass[sigconf,screen]{acmart}
\AtBeginDocument{%
  }

\setcopyright{cc}
\setcctype[4.0]{by}

\copyrightyear{2026}
\acmYear{2026}
\acmDOI{}
\acmConference[arXiv]{arXiv}{preprint}{2026}
\acmBooktitle{arXiv preprint, April, 2026}
\acmISBN{}

\usepackage{url}
\usepackage{verbatim}
\usepackage{graphicx}
\usepackage{tablefootnote}
\usepackage{colortbl}
\usepackage{hyperref}
\usepackage{orcidlink}
\usepackage{listings}
\usepackage{xcolor}
\usepackage[dvipsnames]{xcolor}
\usepackage[framemethod=TikZ]{mdframed}
\usepackage{algorithm}
\usepackage{algpseudocode}
\usepackage{tcolorbox}
\usepackage{listings}
\usepackage{balance}
\usepackage{amsmath}
\DeclareMathOperator*{\argmax}{arg\,max}

\newcommand{\rqone}{\textbf{RQ\textsubscript{1}}. How well does \textsc{FlaXifyer} predict intermittent job failure categories?}

\newcommand{\rqtwo}{\textbf{RQ\textsubscript{2}}. How does model performance evolve as new failure categories are introduced?}

\newcommand{\rqthree}{\textbf{RQ\textsubscript{3}}. Can we determine the most influential log statements to support interpretability?}

\definecolor{light-gray}{gray}{0.95}

\definecolor{medium-gray}{gray}{0.8}

\tcbuselibrary{listings, skins, breakable}

\definecolor{grayframe}{RGB}{100, 100, 100}
\definecolor{graylight}{RGB}{245, 245, 245}

\newtcblisting{logsegment}[1][]{
  listing only,
  colback=graylight,
  colframe=grayframe,
  coltitle=white,
  colbacktitle=grayframe,
  rounded corners,
  arc=2mm,
  boxrule=0.4pt,
  title={#1},
  left=3pt,
  right=3pt,
  top=2pt,
  bottom=2pt,
  listing options={
    basicstyle=\ttfamily\footnotesize,
    breaklines=true,
    showstringspaces=false
  }
}

\begin{document}

\title{Predicting Intermittent Job Failure Categories for Diagnosis Using Few-Shot Fine-Tuned Language Models}

\author{Henri A\"idasso}
\email{henri.aidasso.1@ens.etsmtl.ca}
\orcid{0009-0004-1625-0159}
\affiliation{%
  \institution{École de technologie supérieure}
  \city{Montreal}
  \state{Quebec}
  \country{Canada}
}

\author{Francis Bordeleau}
\email{francis.bordeleau@etsmtl.ca}
\orcid{0000-0001-7727-3902}
\affiliation{%
  \institution{École de technologie supérieure}
  \city{Montreal}
  \state{Quebec}
  \country{Canada}
}

\author{Ali Tizghadam}
\email{ali.tizghadam@telus.com}
\orcid{0000-0002-0898-3094}
\affiliation{%
  \institution{TELUS}
  \city{Toronto}
  \state{Ontario}
  \country{Canada}
}

\renewcommand{\shortauthors}{A\"idasso et al.}

\begin{abstract}

In principle, Continuous Integration (CI) pipeline failures provide valuable feedback to developers on code-related errors. In practice, however, pipeline jobs often fail intermittently due to non-deterministic tests, network outages, infrastructure failures, resource exhaustion, and other reliability issues. These intermittent (\textit{flaky}) job failures lead to substantial inefficiencies: wasted computational resources from repeated reruns and significant diagnosis time that distracts developers from core activities and often requires intervention from specialized teams. Prior work has proposed machine learning techniques to detect intermittent failures, but does not address the subsequent diagnosis challenge. To fill this gap, we introduce \textsc{FlaXifyer}, a few-shot learning approach for predicting intermittent job failure categories using pre-trained language models. \textsc{FlaXifyer} requires only job execution logs and achieves 84.3\% Macro F1 and 92.0\% Top-2 accuracy with just 12 labeled examples per category. We also propose \textsc{LogSift}, an interpretability technique that identifies influential log statements in under one second, reducing review effort by 74.4\% while surfacing relevant failure information in 87\% of cases. Evaluation on 2,458 job failures from TELUS, demonstrates that \textsc{FlaXifyer} and \textsc{LogSift} enable effective automated triage, accelerate failure diagnosis, and pave the way towards the automated resolution of intermittent job failures.

\end{abstract}



\maketitle

\section{Introduction}
\label{sec:introduction}

Central to modern software delivery processes, Continuous Integration and Continuous Deployment (CI/CD) pipelines (also known as builds \cite{aidasso_build_2025}) enable organizations to release high-quality software faster and more frequently to maintain competitive advantage~\cite{hilton_usage_2016, aidasso_build_2025}. Their value, however, hinges on one critical assumption: that build outcomes are deterministic and reliable. Developers depend on this reliability to prioritize work, validate software quality, and, when failures occur, diagnose issues efficiently by examining execution logs~\cite{hilton_trade-offs_2017, hilton_usage_2016}. Unreliable outcomes undermine this entire feedback loop. For our industrial partner TELUS, a leading telecommunications company, dependable builds are not merely desirable but essential for sustaining product quality and team productivity across their networking software products~\cite{aidasso_diagnosis_2025}.

In practice, however, build jobs often fail for reasons unrelated to code changes, acting as false alarms that mislead developers and erode trust in CI~\cite{lampel_when_2021, olewicki_towards_2022, moriconi_automated_2022, aidasso_diagnosis_2025}. These \textit{intermittent job failures} (also termed flaky jobs~\cite{aidasso_diagnosis_2025, moriconi_automated_2022, durieux_empirical_2020}) exhibit non-deterministic behavior: such jobs may fail and subsequently pass upon rerun with no modifications to the source code or CI script~\cite{olewicki_towards_2022, aidasso_efficient_2025}. Root causes vary widely, from transient network issues and resource exhaustion to race conditions, non-deterministic tests, and infrastructure problems~\cite{lampel_when_2021, aidasso_diagnosis_2025}. This diversity poses two challenges: developers waste computational resources repeatedly rerunning jobs expecting they will pass~\cite{aidasso_diagnosis_2025, olewicki_towards_2022}; and when reruns fail to resolve the issue, they face substantial diagnosis overhead, navigating lengthy logs to pinpoint root causes that often require coordination across teams~\cite{lampel_when_2021, olewicki_towards_2022, aidasso_diagnosis_2025}.

Prior work has focused primarily on the first challenge: detecting intermittent job failures from regular ones to reduce wasteful reruns. For this purpose, Lampel et al.~\cite{lampel_when_2021} trained a binary classifier using job telemetry data from Mozilla's CI service. Since telemetry is not always available, Olewicki et al.~\cite{olewicki_towards_2022} proposed an approach using TF-IDF~\cite{ramos_using_2003} representation of job logs combined with rerun metrics. Moriconi et al.~\cite{moriconi_automated_2022} explored knowledge graphs with limited results, while substantial work has focused specifically on flaky tests~\cite{lam_idflakies_2019, parry_flake_2020, bell_deflaker_2018, akli_flakycat_2023, alshammari_flakeflagger_2021, fatima_flakyfix_2024, rahman_flakesync_2024}, one cause of intermittent job failures. However, flaky tests account for only a small proportion of such failures~\cite{aidasso_diagnosis_2025}, whose root causes are far more diverse~\cite{aidasso_diagnosis_2025, lampel_when_2021, durieux_empirical_2020}, and any build job—not just test jobs—can fail intermittently~\cite{lampel_when_2021, olewicki_towards_2022, aidasso_diagnosis_2025}.

The second challenge, that of diagnosing intermittent job failures, remains largely unaddressed \cite{aidasso_diagnosis_2025}. While some transient issues (e.g., network outage) resolve upon job rerun, many failure categories, including infrastructure problems, DNS resolution errors, and credential issues, require substantial diagnosis effort~\cite{lampel_when_2021, olewicki_towards_2022, aidasso_diagnosis_2025}. This overhead distracts developers from core activities and delays software releases \cite{aidasso_diagnosis_2025}. At TELUS, the diagnosis of such failures often requires coordination across network, infrastructure, and application teams, extending resolution time to several days~\cite{aidasso_diagnosis_2025}.

There is therefore a pressing need for automated methods to categorize intermittent job failures and support diagnosis~\cite{aidasso_diagnosis_2025, lampel_when_2021}. Our prior work~\cite{aidasso_diagnosis_2025} contributed a catalogue of 46 failure categories at TELUS and an open-source automated labeling tool \cite{aidasso_flakeranker_2025}. However, this regex-based tool lacks semantic understanding, cannot generalize to unseen log patterns, and requires manual, error-prone maintenance, limiting its practical applicability.

To bridge this gap, we introduce \textsc{FlaXifyer}, an approach for predicting intermittent (\textit{flaky}) job failure categories using few-shot fine-tuned language models. \textsc{FlaXifyer} requires only job execution logs to predict failure categories, making it practical for deployment. Given the heterogeneous nature of job logs~\cite{brandt_logchunks_2020}, we compare two encoder types: BGE~\cite{xiao_c-pack_2024}, a general-purpose text encoder, and CodeBERT~\cite{feng_codebert_2020}, a code-aware model. Both are fine-tuned using a few-shot learning technique~\cite{tunstall_efficient_2022}, proven effective under limited labeled data. We evaluate \textsc{FlaXifyer} on a labeled dataset of 2,458 intermittent job failures from industrial projects at TELUS~\cite{aidasso_diagnosis_2025} and propose \textsc{LogSift}, a technique for identifying influential log statements to support interpretability. This study addresses the following research questions (RQs):

\textbf{\rqone} Evaluated across different shot settings (2--16) and encoder types (BGE vs. CodeBERT), \textsc{FlaXifyer} achieves 84.3\% Macro F1 and 92.0\% Top-2 accuracy with BGE at 12 shots, outperforming CodeBERT by 8.1 percentage points. Performance plateaus around 10--12 shots, and general-purpose encoders prove more effective than code-aware models for heterogeneous job logs.

\textbf{\rqtwo} We assess \textsc{FlaXifyer}'s scalability by incrementally extending from 8 core categories to 13. Macro F1 decreases modestly from 92.5\% to 84.3\%, with most core categories remaining stable. Per-class F1 ranges from 68.4\% to 98.8\%, revealing categories that are easy (e.g., \texttt{runner pod waiting timeout}) and difficult (e.g., \texttt{host resolution failure}) to classify.

\textbf{\rqthree} We evaluate \textsc{LogSift}, our interpretability technique for identifying influential log statements. \textsc{LogSift} achieves 74.4\% mean reduction ratio with 63.1\% of its outputs contain 30 lines or fewer. Qualitative analysis confirms that 87\% of identified log segments are relevant for diagnosis, surfacing actionable failure information in under one second.

In summary, this paper makes the following contributions:

\begin{itemize}
\item \textbf{\textsc{FlaXifyer}}, a few-shot learning approach for predicting intermittent job failure categories using pre-trained language models, achieving 84.3\% F1-score.
\item \textbf{\textsc{LogSift}}, an interpretability technique for identifying influential log statements to support diagnosis.
\item \textbf{Replication package} \cite{aidasso_artifact_2026}, including implementation scripts and intermediary results, for verification and replication.
\end{itemize}

The remainder of this paper is organized as follows. Section~\ref{sec:background} presents background and related work. Section~\ref{sec:flaxifyer} details \textsc{FlaXifyer}, our approach for predicting intermittent job failure categories. It further presents \textsc{LogSift}, our interpretability technique to identify influential log statements. Section~\ref{sec:study_design} describes the study design, and Section~\ref{sec:results} reports experimental results. Section~\ref{sec:discussion} discusses findings and implications. Finally, Section~\ref{sec:threats} presents threats to validity, and Section~\ref{sec:conclusion} concludes the paper with future directions.

\section{Background and Related Work}
\label{sec:background}

\subsection{CI/CD and Intermittent Job Failures}

Continuous Integration and Continuous Deployment (CI/CD) are software engineering practices that automate the building, testing, and deployment of code changes, enabling organizations to deliver high-quality software rapidly and frequently~\cite{humble_continuous_2010, hilton_usage_2016}. In fact, developers are encouraged to commit small, incremental code changes frequently to the central code repository. Each commit triggers a CI/CD pipeline comprising multiple jobs, such as code compilation, static code analysis, testing, packaging, and deployment, that execute automatically to deliver a new software version~\cite{aidasso_build_2025}.

Each job execution follows a structured sequence of steps, producing heterogeneous logs. Common steps, such as runner environment setup, repository cloning, dependency installation, artifact caching, and cleanup, are shared across virtually all jobs, while job-specific steps execute the commands defined for that particular task (e.g., compilation, testing, image building, deployment). Each step produces textual logs, resulting in extensive and highly heterogeneous job execution logs that mix natural language, shell commands and outputs, code snippets, error stack traces, package installation traces, and tool-specific outputs~\cite{brandt_logchunks_2020}. Developers rely on these job logs for traceability and failure diagnosis, though their verbosity and heterogeneity make manual inspection challenging~\cite{aidasso_diagnosis_2025, olewicki_towards_2022}.

Unlike regular code-related failures, which provide deterministic feedback on software defects or quality issues, intermittent job failures typically arise from factors external to the submitted code. These factors are very diverse, including network outages, resource exhaustion, infrastructure instabilities, authentication errors, and timeouts~\cite{lampel_when_2021, aidasso_diagnosis_2025, durieux_empirical_2020}. Also referred to as flaky or brown builds~\cite{olewicki_towards_2022}, intermittent job failures exhibit non-deterministic outcomes, potentially passing upon rerun without any modification to the source code or CI script. The unreliable nature of these failures erodes developer trust in CI, prompting repeated reruns that waste computing resources and prolonged diagnosis efforts that distract engineers from core development activities~\cite{lampel_when_2021, olewicki_towards_2022, aidasso_diagnosis_2025}. Significant rates of intermittent job failures have been reported across organizations, including at Mozilla~\cite{lampel_when_2021} and Ubisoft~\cite{olewicki_towards_2022}, highlighting an important and widespread industry challenge.

\paragraph{Industrial Context.} TELUS is a leading Canadian telecommunications company developing software products for self-driving networks, including the TELUS Intelligent Network Analytics and Automation (TINAA) ecosystem~\cite{telus_telus_2021}. TINAA comprises microservice-based applications serving millions of end-users across critical sectors, such as city-wide telecommunications, internet, television, telehealth, and smart agriculture. In this context, CI/CD reliability is paramount to ensure safe and reliable software delivery. 

However, the problem of intermittent job failures is particularly acute at TELUS, where CI jobs execute on an in-house GitLab CI setup including Kubernetes (k8s) pods and multiple distributed services: container registries, cloud environments, security scanners, and artifact storage. This distributed architecture introduces numerous failure points, with intermittent failures accounting for 25\% of all job failures~\cite{aidasso_diagnosis_2025}. The associated reruns strain already limited CI resources, while prolonged diagnosis efforts reduce developer productivity and incur significant costs~\cite{aidasso_diagnosis_2025, aidasso_efficient_2025}. Compounding the problem, the root causes of these issues often span multiple domains, requiring involvement from network, infrastructure, and application teams and extending resolution to several days~\cite{aidasso_diagnosis_2025}.

\paragraph{Motivation for Automated Failure Categorization.}
Diagnosing intermittent job failures is particularly time-consuming, as root causes often lie outside developers' domain of responsibility~\cite{aidasso_diagnosis_2025, olewicki_towards_2022}. At TELUS, the diagnosis of intermittent failures is thus further complicated by organizational complexity. For instance, failures involving infrastructure-managed resources (e.g., deployment tokens, network devices, container registry) require coordination across multiple teams. Some logs even contain instructions such as \textit{``\dots please contact infrastructure team\dots''} buried within thousands of lines that developers must review manually. Automated failure categorization addresses these challenges in three ways: (1) helping developers quickly understand the nature of the issue they face, (2) enabling automated triage by directing failures to the appropriate team, and (3) serving as a prerequisite for category-specific automated repair strategies. Complementing categorization with relevant log segments further accelerates diagnosis by surfacing failure-specific information that can accompany triage notifications, reducing the time teams spend identifying root causes for resolution.

\subsection{Related Work}

\subsubsection{Intermittent Job Failure Detection}

Techniques for detecting intermittent job failures have been developed to distinguish them from regular code-related failures, aiming to reduce the costly manual reruns that developers often resort to. Lampel et al.~\cite{lampel_when_2021} proposed, to our knowledge, the first approach at Mozilla. They leveraged job telemetry data (e.g., CPU load, runtime, machine name) from 2 million jobs to train tree-based classifiers, achieving 73\% average precision. However, this approach requires costly manual labeling and relies on telemetry data that is difficult or impossible to obtain due to restricted access and multi-tenancy noise in CI processes using public or shared infrastructure~\cite{olewicki_towards_2022}. To address these limitations, Olewicki et al.~\cite{olewicki_towards_2022} introduced an approach that uses TF-IDF~\cite{ramos_using_2003} representation of job logs, combined with job rerun-related metrics, to train XGBoost classifiers. 
Moriconi et al.~\cite{moriconi_automated_2022} also explored knowledge graphs for detection with promising early-stage results.
Another line of research has addressed flaky tests~\cite{lam_idflakies_2019, bell_deflaker_2018, akli_flakycat_2023, alshammari_flakeflagger_2021, fatima_flakyfix_2024, rahman_flakesync_2024}, one source of intermittent job failures. However, flaky tests constitute only a fraction (about 4\% at TELUS \cite{aidasso_diagnosis_2025}) of such failures, which can affect any job type and stem from far more diverse causes~\cite{aidasso_diagnosis_2025, lampel_when_2021}.

While these detection approaches help reduce wasteful reruns, they do not address the substantial overhead of \textit{diagnosing} intermittent failures. This is particularly challenging at TELUS, where identifying the appropriate team (network, infrastructure, or application) to handle a failure can itself consume significant time and further delay resolution of pipeline reliability issues~\cite{aidasso_diagnosis_2025, aidasso_illusion_2025}.

\subsubsection{Intermittent Job Failure Categories}

In prior work~\cite{aidasso_diagnosis_2025}, we analyzed intermittent job failures at TELUS and introduced a comprehensive catalogue of 46 failure categories, along with a regex-based labeling tool~\cite{aidasso_flakeranker_2025}. Using Recency-Frequency-Monetary (RFM) analysis~\cite{funatsu_data_2011}, we identified 14 priority categories that warrant attention for automated diagnosis support, including:

\begin{itemize}
    \item \textbf{misconfigured env variable}: failures caused by misspelled, missing, or invalid CI/CD environment variables;
    \item \textbf{job execution timeout}: failures due to jobs exceeding predefined execution time limits;
    \item \textbf{flaky ui test}: flaky test failures in UI components due to timing issues, element visibility, or async rendering;
    \item \textbf{git transient error}: temporary issues during \textit{git} operations, typically caused by network or server instability;
    \item \textbf{host resolution failure}: DNS resolution failures when attempting to connect to remote hosts or services.
\end{itemize}

While essential for obtaining labeled data, the existing regex-based tool \cite{aidasso_flakeranker_2025} has significant limitations for automated categorization. First, handcrafted regex patterns do not account for semantic or structural context and can miss even slight log variations. Second, maintaining the tool is tedious and error-prone, requiring manual updates whenever log patterns change. TELUS engineers emphasized the need for an automated solution that scales reliably across projects and remains easy to maintain.

Overall, existing approaches focus on {detecting} intermittent failures to reduce wasteful reruns, but leave the diagnosis challenge largely unaddressed.  This study presents the first approach for \textit{predicting intermittent job failure categories} to support practitioners in the diagnosis and repair of such failures.

\section{\textsc{FlaXifyer}}
\label{sec:flaxifyer}

\begin{figure*}[!t]
\centering
\includegraphics[width=.85\textwidth]{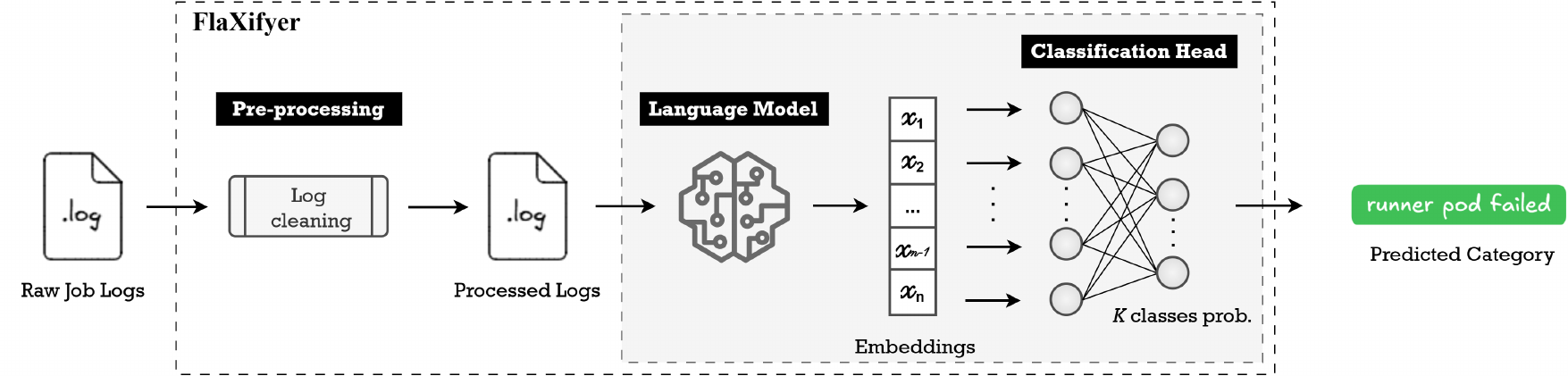}
\caption{Predicting intermittent job failure categories from job logs using a language model and classification head.}
\label{fig:flaxifyer_overview}
\end{figure*}

\subsection{Overview}

\textsc{FlaXifyer} addresses the diagnosis challenge through \textit{multi-class classification} of intermittent job failure categories. We name our approach \textsc{FlaXifyer} to emphasize its ability to classify intermittent (a.k.a \textit{flaky}) job failures. Unlike prior works on flaky tests~\cite{akli_flakycat_2023, fatima_flakify_2023, fatima_flakyfix_2024}, which address that specific subset of job-related issues, \textsc{FlaXifyer} targets the broader challenge of categorizing diverse flaky job failures across the CI/CD pipeline. Given a job's execution logs, it predicts the failure's root cause category (e.g., \texttt{runner pod failure}, \texttt{docker daemon connection failure}, \texttt{job exec timeout}), enabling developers to quickly route issues to the appropriate team or, in the future, apply category-specific automated repair strategies.

Figure~\ref{fig:flaxifyer_overview} illustrates the \textsc{FlaXifyer} pipeline, which comprises three stages: (1) pre-processing raw job logs to reduce noise and size, (2) generating numerical embeddings using a fine-tuned language model (specifically, a text encoder), and (3) predicting the failure category by classifying the embeddings produced in stage (2). \textsc{FlaXifyer} is designed to be practical. It requires only job execution logs as input and leverages few-shot learning to minimize the manual labeling effort for training. At prediction time, given a raw log file, it outputs the predicted failure category $c$ as:
\begin{equation}
    c = C(E(P(\text{log})))
\end{equation}
where $P(\cdot)$ denotes the log pre-processing function, $E(\cdot)$ the fine-tuned encoder, and $C(\cdot)$ the classification head.

The following subsections detail each stage of implementation.

\subsection{Log Pre-processing}

Job logs are inherently heterogeneous, verbose, and partially repetitive~\cite{brandt_logchunks_2020}, making them susceptible to the curse of dimensionality when represented as vectors. Moreover, the input length of language models used in the embedding stage is limited to 512 tokens. Pre-processing is therefore essential to reduce log size while preserving information critical for identifying the failure category.

We apply a set of pre-processing rules adapted from prior work on job log analysis~\cite{aidasso_efficient_2025, olewicki_towards_2022, brandt_logchunks_2020}. These rules were refined through iterative manual inspection of 100 randomly sampled logs to ensure that variable elements (e.g., URLs, IDs) are properly abstracted and noisy elements (e.g., special characters) are removed, while preserving key failure-related information.

\begin{itemize}
    \item \textbf{Rule 1.} URLs, file paths, directory paths, durations, and versions are replaced by constant tokens (\texttt{<URL>}, \texttt{<FILEPATH>}, \texttt{<DIRPATH>}, \texttt{<DURATION>}, \texttt{<VERSION>}).
    \item \textbf{Rule 2.} Identifiers (alphanumeric sequences containing at least one letter and one digit) are replaced by \texttt{<ID>}.
    \item \textbf{Rule 3.} Non-alphanumeric characters are removed.
    \item \textbf{Rule 4.} Numbers are removed, except HTTP status codes and exit codes, which are highly relevant for diagnosis.
    \item \textbf{Rule 5.} Trailing single letters are removed.
    \item \textbf{Rule 6.} Whitespace and blank lines are removed.
    \item \textbf{Rule 7.} Duplicate lines are removed.
\end{itemize}

Overall, these rules reduce raw log size by an average of 61\%, consistent with prior work~\cite{aidasso_efficient_2025, olewicki_towards_2022}.

\subsection{Few-Shot Fine-Tuning}
\label{sec:finetuning}

The first step of \textsc{FlaXifyer}’s training process involves few-shot fine-tuning of pre-trained embedding models to produce representations that clearly separate intermittent failure categories. Prior work on intermittent job failure detection~\cite{aidasso_efficient_2025} has shown that fine-tuned embedding models significantly outperform TF--IDF representations for binary job log classification, motivating our choice of embedding models for this multi-class classification task. 

To this end, we use the SetFit framework~\cite{tunstall_efficient_2022}, which enables efficient, prompt-free fine-tuning using only a small number of labeled examples. This capability is essential, as manually annotating job failures with their root-cause categories is costly and time-consuming. While our regex-based labeling tool~\cite{aidasso_flakeranker_2025} can generate an initial labeled dataset, it lacks semantic understanding, does not generalize well to log variations, and requires substantial (error-prone) manual maintenance to support new failure categories. Few-shot fine-tuning addresses these limitations by learning semantic representations from minimal labeled data.

Specifically, SetFit fine-tunes the pre-trained embedding model using contrastive learning~\cite{koch_siamese_2015} to learn a discriminative embedding space. It generates a dataset of triplets from the training set, including (1) \textit{positive triplets} $\{(x_i, x_j, 1)\}$, where log pairs are sampled from the \textit{same} failure category; and (2) \textit{negative triplets} $\{(x_i, x_j, 0)\}$, where pairs are sampled from \textit{different} categories. Using a cosine similarity loss, the encoder learns to position logs from the same category closer together in the embedding space while pushing logs from different categories farther apart. This contrastive objective is well-suited for multi-class problems, as it learns pairwise relationships that generalize across all categories.

Once fine-tuned, the encoder generates embeddings for all training logs, which are then used to train the classifier (Section~\ref{sec:classification}).

\subsection{Classification Head}
\label{sec:classification}

The second training stage fits a classifier on the embeddings produced by the fine-tuned encoder. We use Logistic Regression with multinomial loss as the classification head, mapping 768-dimensional embeddings to failure categories. Because contrastive fine-tuning produces well-separated embeddings, this classifier is both fast to train and effective under limited labeled data, consistent with prior work~\cite{aidasso_efficient_2025} and SetFit's recommended configuration~\cite{tunstall_efficient_2022}. The classifier outputs a probability distribution over the $K$ failure categories.

At inference time, given a raw job log, \textsc{FlaXifyer} first applies log pre-processing $P$, then encodes the processed log using the fine-tuned encoder $E$, and finally predicts a failure category via the trained classifier $C$. Formally, the predicted category $c$ is given by:
\begin{equation}
    c = \argmax_{k \in \{1, \dots, K\}} \; \bigl[C\bigl(E(P(\text{log}))\bigr)\bigr]_k
\end{equation}

\subsection{Interpretability: Identifying Influential Log Statements}
\label{sec:interpretability}

\subsubsection{Motivation}

While accurate prediction of failure categories supports diagnosis, practitioners also need to understand \textit{why} a particular category was predicted. Interpretability is essential for building trust in automated classification systems and enabling developers to verify predictions against their domain knowledge~\cite{ribeiro_why_2016}. To address this need, we introduce \textsc{LogSift}, a technique for identifying the log statements most influential to the classifier's decision.

\subsubsection{Approach}

Our approach draws inspiration from binary search \cite{knuth_art_1998}, recursively partitioning the logs to isolate the minimal segments that preserve the predicted category. The intuition is that if a log segment contains sufficient information for the classification pipeline to predict the same category as the full log, then the influential statements lie within that segment. Algorithm~\ref{alg:logsift} describes the procedure. Given the statements $L$ of a job log and the original category $c$ predicted from the full log, we recursively bisect $L$ into top and bottom halves. When the number of statements is odd, the bottom half receives the extra statement. For each half, we query the classification pipeline $f$: if the predicted category matches $c$, we continue the search within that half; otherwise, we backtrack. When both halves independently predict $c$, influential information is distributed across both segments, and we recursively search each half, aggregating the results. The recursion terminates when the segment size falls below a minimum threshold $\tau$ (set to 2 statements to ensure sufficient context) or when neither half preserves the original prediction. In the latter case, the current segment is the minimal chunk containing the influential statements. For practicality, we return a list of $(\texttt{start}, \texttt{end})$ index ranges identifying the influential statements, which can then be used to highlight or extract the corresponding log segments for developer review. The proposed algorithm has logarithmic complexity $O(\log n)$ in the number of log statements $n$ when influential information is localized, and linear complexity $O(n)$ in the worst case when information is distributed throughout the log. In practice, important failure-related statements tend to cluster near error messages, yielding efficient execution even for verbose logs.

\begin{algorithm}[t]
\caption{\textsc{LogSift}: Influential Log Statements Identification}
\label{alg:logsift}
\begin{algorithmic}[1]
\Require $L$: Raw job log as a list of log statements.
\Require $f$: \textsc{FlaXifyer} classification pipeline
\Require $\tau$: minimum segment size (default: 2)
\Ensure $R$: list of $(\text{start}, \text{end})$ index ranges
\Function{LogSift}{$L$, $f$, $\tau$}
    \State $c \gets f(L)$ \Comment{Original predicted category}
    \State \Return \Call{FindInfluential}{$L$, $c$, $f$, $\tau$, $0$}
\EndFunction
\Function{FindInfluential}{$L$, $c$, $f$, $\tau$, $\text{offset}$}
    \If{$|L| \leq \tau$}
        \State \Return $[(\text{offset}, \; \text{offset} + |L| - 1)]$
    \EndIf
    \State $m \gets \lfloor |L| / 2 \rfloor$
    \State $L_{top} \gets L[0 : m]$
    \State $L_{bot} \gets L[m : |L|]$ \Comment{Larger half if $|L|$ is odd}
    \State $match_{top} \gets f(L_{top}) = c$
    \State $match_{bot} \gets f(L_{bot}) = c$
    \If{$match_{top}$ \textbf{and} $match_{bot}$}
        \State \Return \Call{FindInfluential}{$L_{top}$, $c$, $f$, $\tau$, $\text{offset}$}
        \State \hspace{\algorithmicindent} $\cup$ \Call{FindInfluential}{$L_{bot}$, $c$, $f$, $\tau$, $\text{offset} + m$}
    \ElsIf{$match_{top}$}
        \State \Return \Call{FindInfluential}{$L_{top}$, $c$, $f$, $\tau$, $\text{offset}$}
    \ElsIf{$match_{bot}$}
        \State \Return \Call{FindInfluential}{$L_{bot}$, $c$, $f$, $\tau$, $\text{offset} + m$}
    \Else
        \State \Return $[(\text{offset}, \; \text{offset} + |L| - 1)]$ \Comment{Minimal segment preserving $c$}
    \EndIf
\EndFunction
\end{algorithmic}
\end{algorithm}

\subsubsection{Practical Benefits}

This interpretability technique offers several advantages for practitioners. First, it highlights the specific log statements that drove the prediction, allowing developers to quickly verify whether the classification is reasonable. Second, it reduces the cognitive load of manually inspecting lengthy logs by directing attention to the most relevant segments, which include valuable specific information to accelerate the resolution process. Third, when the model makes errors, examining the influential segments can reveal whether the misclassification stems from ambiguous log patterns or model limitations, informing future improvements.

\section{Study Design}
\label{sec:study_design}

\subsection{Research Questions}
\label{sec:rqs}

The primary goal of this study is to investigate the effectiveness of \textsc{FlaXifyer} in predicting intermittent job failure categories, its scalability as new categories emerge, and its interpretability for practical use. Specifically, we address the following RQs:

\vspace{0.5em}
\noindent\textbf{\rqone} 
Identifying the optimal shot setting and encoder type is essential for practical deployment. On the one hand, identifying the optimal number of shots determines how many examples must be manually labeled to train an effective classifier and, importantly, to extend it with new failure categories in the future. On the other hand, job logs are heterogeneous, containing natural language messages, code fragments, stack traces, and tool-specific outputs. Whether a general-purpose or code-aware pre-trained embedding model better captures this heterogeneity remains unclear. We therefore evaluate our approach, comparing two encoders: BGE (general-purpose) and CodeBERT (code-aware), across shot settings ranging from 2 to 16 to determine the most effective configuration for predicting intermittent job failure categories using the job logs.

\vspace{0.5em}
\noindent\textbf{\rqtwo}
In practice, new failure categories emerge as CI/CD environments evolve. At TELUS, infrastructure changes, new tools are adopted, and previously rare issues become prevalent~\cite{aidasso_diagnosis_2025}. A practical classification model must accommodate new categories without significant degradation on existing ones. This question evaluates \textsc{FlaXifyer}'s scalability by incrementally extending the classifier from 8 core categories to 13, and examines per-class performance to identify which categories are easy or difficult to classify. These findings will inform practitioners about the expected trade-offs when extending the model and highlight categories that may require additional attention.

\vspace{0.5em}
\noindent\textbf{\rqthree} Accurate predictions alone are insufficient for practical adoption: developers need to understand \textit{why} a particular category was predicted to verify the classification, build trust in the system~\cite{ribeiro_why_2016}, and act on it efficiently. This is particularly challenging given the size of job logs, which range from 13 to 27,532 lines in our dataset (mean: 405). Manually sifting through hundreds or thousands of lines to locate failure-relevant information is impractical, especially when developers face multiple failures daily. Effective interpretability addresses both challenges simultaneously. By highlighting the log statements that drive a prediction, developers can (1) verify the classification \textit{and} (2) immediately access the failure-specific details needed for repair, such as the URL for a \texttt{host resolution failure} or the variable name for a \texttt{misconfigured env variable}. To this end, we evaluate \textsc{LogSift} (Section~\ref{sec:interpretability}), assessing whether it identifies meaningful failure indicators while substantially reducing log review effort.

\subsection{Dataset}
\label{sec:dataset}

\subsubsection{Data Source}

We build upon the dataset from our prior study on intermittent job failure categories at TELUS~\cite{aidasso_diagnosis_2025}. This dataset comprises 4,511 labeled intermittent job failures extracted from 80 industrial GitLab projects spanning 6.5 years of build history, from February 2018 to July 2024. The projects vary in size, maturity, and purpose (including software-defined networks, data analytics, APIs, and web portals) and cover 13 programming languages. Each instance is labeled with one of 46 root cause categories identified through semi-automated labeling using a regex-based tool, achieving 91\% precision on an unseen representative sample~\cite{aidasso_diagnosis_2025}.

\subsubsection{Category Selection}

Multi-class classification with 46 categories is impractical. It presents challenges related to class imbalance and label sparsity, where many categories contain insufficient examples for effective model training. To address this, we focus on the 14 priority categories identified in our prior work through RFM analysis~\cite{aidasso_diagnosis_2025}. These categories represent the most recent, frequent, and costly failures, i.e. those causing the greatest waste of developer time and CI resources at TELUS. Moreover, they alone account for more than half (55\%) of the labeled 4,511 intermittent job failures. Of the 14 priority categories, we exclude \texttt{cloud\_token\_limit\_exceeded} (39 examples) due to insufficient samples for few-shot training at higher shot settings, as detailed in Section~\ref{sec:experimental_setup}. Table~\ref{tab:categories} presents the 13 categories selected, ranked by RFM priority.

\begin{table}[t]
\centering
\caption{Selected priority failure categories ranked by RFM priority: top (rank 1--2), high (rank 3--8), medium (rank 9--13).}
\label{tab:categories}
\begin{tabular}{clrr}
\toprule
\textbf{Rank} & \textbf{Category} & \textbf{Count} & \textbf{\%} \\
\midrule
1 & misconfigured\_env\_variable & 673 & 27.4 \\
2 & job\_execution\_timeout & 306 & 12.4 \\
3 & dependency\_installation\_failure & 124 & 5.0 \\
4 & runner\_pod\_waiting\_timeout & 199 & 8.1 \\
5 & api\_gateway\_deployment\_error & 161 & 6.5 \\
6 & container\_registry\_server\_error & 213 & 8.7 \\
7 & git\_transient\_error & 113 & 4.6 \\
8 & flaky\_ui\_test & 179 & 7.3 \\
9 & external\_file\_invalid\_format & 125 & 5.1 \\
10 & host\_resolution\_failure & 91 & 3.7 \\
11 & runner\_image\_pull\_failure & 99 & 4.0 \\
12 & remote\_call\_timeout & 76 & 3.1 \\
13 & helm\_resource\_error & 99 & 4.0 \\
\midrule
& \textbf{Total} & \textbf{2,458} & \textbf{100.0} \\
\bottomrule
\end{tabular}
\end{table}

\subsubsection{Data Labeling and Limitations}

Labels were obtained using the regex-based tool~\cite{aidasso_flakeranker_2025} from our prior work~\cite{aidasso_diagnosis_2025}. While effective for dataset construction, this tool has inherent limitations: regex patterns lack semantic understanding and fail to capture log variations, resulting in incomplete coverage. Manual review revealed that the tool missed approximately 15\% of categorizable failures due to minor variations (e.g., \texttt{out-of-memory} vs. \texttt{insufficient space}). Furthermore, extending the tool to new categories requires substantial manual effort to craft and maintain patterns. These limitations motivate our machine learning approach and the use of few-shot learning, which enables high-quality manual labeling of a small number of examples rather than relying on erroneous automated labeling at scale.

\subsection{Experimental Setup}
\label{sec:experimental_setup}

We adopt the evaluation methodology from our prior work on binary failure classification~\cite{aidasso_efficient_2025}, extending it to multi-class classification across intermittent failure categories.

\subsubsection{Few-Shot Training Configuration}

Following SetFit~\cite{tunstall_efficient_2022} requirements, the training set $TR$ is constructed by randomly sampling $N$ log examples for each of the $K$ failure categories, yielding a training set of size $N \times K$. In line with recommendations for few-shot text classification, we use Logistic Regression as the classification head~\cite{tunstall_efficient_2022}. The choice of the underlying embedding model is treated as a separate design decision and is investigated in RQ\textsubscript{1}, where we analyze its impact on classification performance.

\subsubsection{Cross-Validation Pipeline}

To evaluate model performance under a fixed few-shot regime, we use Monte Carlo Cross-Validation (MCCV)~\cite{simon_resampling_2007}, also known as repeated random subsampling. This method provides robust performance estimates by averaging evaluation results over multiple random data splits, thereby reducing variance from the random selection of training examples. Besides, compared to traditional $k$-fold cross-validation, MCCV has been shown to provide more reliable estimates in low-resource settings~\cite{shan_monte_2022}.

Each MCCV iteration $i$ proceeds as follows. First, the input dataset $D$ is stratified and randomly split into a learning set $L_i$ (25\%), a validation set $V_i$ (25\%), and a test set $T_i$ (50\%). From $L_i$, we then sample $N$ labeled examples per failure category to form the few-shot training set $TR_i$. Model hyperparameters are optimized using $V_i$, and the resulting best model is evaluated on $T_i$ to obtain the model performance estimate $\varepsilon_i$. Allocating a larger portion of the dataset $D$ to the test set $T_i$ improves the reliability of performance estimates, while the learning set remains sufficiently large to support few-shot sampling. We set $i_{\max}=30$ independent iterations, and the final model performance is computed as:

\begin{equation}
    \bar{\varepsilon} = \frac{1}{i_{max}} \sum_{i=1}^{i_{max}} \varepsilon_i
\end{equation}

\textbf{Category exclusion.} For reliable $N$-shot sampling across MCCV iterations, each category must contain sufficient examples in the learning set. With a 25\% learning split, categories require at least $4N$ total examples to consistently support $N$-shot training. Among the 14 previously identified priority categories~\cite{aidasso_diagnosis_2025}, the failure category \texttt{cloud\_token\_limit\_exceeded} (39 examples) does not meet this threshold at higher shot settings evaluated in RQ\textsubscript{1} (39 $<$ 64 examples required for 16-shots) and is therefore excluded. The 13 remaining categories contain between 76 and 673 examples each (Table~\ref{tab:categories}), ensuring adequate support for the experiments.

\subsubsection{Hyperparameter Optimization}

We tune model training hyperparameters using Optuna~\cite{akiba_optuna_2019} with 5 trials per model configuration, based on a recent study and fine-tuning guidelines~\cite{pareja_unveiling_2024, aidasso_efficient_2025}. During each trial, hyperparameter values are sampled from the following hyperparameter space:

\begin{itemize}
    \item \textbf{Body learning rate}: $10^{-6}$ to $10^{-3}$
    \item \textbf{Number of epochs}: 1 or 2
    \item \textbf{Batch size}: 2, 4, or 8
    \item \textbf{Max iter}: 50 to 300, by 50
\end{itemize}

The hyperparameter values yielding the highest performance on the validation set are selected for final evaluation on the test set.

\subsubsection{Computational Environment}
Model training was performed on NVIDIA A100 GPUs (40GB), with training time ranging from under 1 minute to 70 minutes per iteration (median: 3.3 minutes). \textsc{LogSift} evaluation and RQs analysis were executed on CPU nodes with 32GB RAM. All experiments used Python 3.11.

\subsection{Evaluation Metrics}
\label{sec:metrics}

We evaluate \textsc{FlaXifyer} using metrics suited for multi-class classification with imbalanced class distributions, since our category frequencies range from 76 to 673 examples (Table~\ref{tab:categories}). All metrics are aggregated across the $i_{max}=30$ MCCV iterations and reported as mean $\pm$ standard deviation.

\subsubsection{Primary Metric}
We use \textbf{Macro F1-score} as the primary evaluation metric. Macro F1 is the unweighted mean of per-class F1-scores, computed as $\text{Macro F1} = \frac{1}{K} \sum_{k=1}^{K} F1_k$, where $K$ denotes the number of categories and $F1_k = \frac{2\,\mathrm{TP}}{2\,\mathrm{TP} + \mathrm{FP} + \mathrm{FN}}$ is the F1-score for category $k$. By treating all categories equally regardless of frequency, Macro F1 penalizes models that perform well on frequent categories but poorly on rare ones. This aligns with our practical objective of supporting diagnosis as even less frequent failures can incur substantial diagnosis cost when they occur~\cite{aidasso_diagnosis_2025}.

\subsubsection{Secondary Metrics}

To provide a comprehensive view of model performance, we report traditional metrics including \textit{precision} $(\frac{TP}{TP + FP})$ and \textit{recall} $(\frac{TP}{TP + FN})$, along with the following: 

\textbf{Matthews Correlation Coefficient (MCC).} MCC provides a balanced measure of classification quality that accounts for all entries in the confusion matrix~\cite{matthews_comparison_1975}. As a result, it produces high scores only when the model performs well across all categories. This score ranges from $-1$ (complete disagreement) to $+1$ (perfect prediction), with 0 indicating random performance.

\textbf{Top-$k$ Accuracy.} The proportion of instances where the correct category appears among the model's top $k$ predictions. We report Top-2 and Top-3 accuracy to assess \textsc{FlaXifyer}'s practical utility in scenarios where presenting multiple candidate categories is acceptable. Even when the top prediction is incorrect, suggesting the correct category among the top few options can substantially reduce diagnosis time by narrowing the scope of investigation.

\section{Experimental Results}
\label{sec:results}

\begin{table*}[t]
\caption{Model performance across shot settings for two encoder families.
Values are reported as mean\textsubscript{std} over seeds (\%).}
\label{tab:setting_comparison}
\centering
\resizebox{\textwidth}{!}{
\begin{tabular}{lcccccccccccccc}
\toprule
 & \multicolumn{7}{c}{\textbf{CodeBERT}} & \multicolumn{7}{c}{\textbf{BGE}} \\
\cmidrule(lr){2-8} \cmidrule(lr){9-15}
\textbf{\#shots}
& f1 & mcc & precision & recall & top-1 & top-2 & top-3
& f1 & mcc & precision & recall & top-1 & top-2 & top-3 \\
\midrule
2  
& 52.9\textsubscript{5.4} & 49.9\textsubscript{6.1} & 57.3\textsubscript{4.5} & 55.6\textsubscript{5.1} & 53.7\textsubscript{6.2} & 64.1\textsubscript{6.3} & 70.1\textsubscript{6.2}
& 66.1\textsubscript{6.2} & 64.7\textsubscript{6.4} & 69.1\textsubscript{6.5} & 69.6\textsubscript{5.2} & 67.5\textsubscript{6.3} & 76.1\textsubscript{6.3} & 80.7\textsubscript{6.3} \\

4  
& 62.3\textsubscript{3.9} & 60.3\textsubscript{4.5} & 64.0\textsubscript{3.9} & 65.6\textsubscript{3.7} & 63.6\textsubscript{4.6} & 73.2\textsubscript{4.8} & 79.1\textsubscript{4.7}
& 75.3\textsubscript{4.2} & 74.4\textsubscript{4.9} & 77.1\textsubscript{4.1} & 78.5\textsubscript{3.4} & 76.5\textsubscript{4.8} & 83.5\textsubscript{5.0} & 88.1\textsubscript{4.8} \\

8  
& 71.2\textsubscript{3.5} & 69.4\textsubscript{4.9} & 71.5\textsubscript{3.5} & 74.0\textsubscript{2.7} & 72.3\textsubscript{4.9} & 81.7\textsubscript{4.3} & 86.2\textsubscript{4.1}
& 82.2\textsubscript{3.7} & 81.5\textsubscript{4.8} & 83.6\textsubscript{3.8} & 84.5\textsubscript{2.8} & 83.1\textsubscript{4.7} & 89.9\textsubscript{3.8} & 93.3\textsubscript{3.7} \\

10 
& 73.8\textsubscript{2.9} & 72.3\textsubscript{4.2} & 73.7\textsubscript{3.0} & 76.6\textsubscript{2.3} & 75.0\textsubscript{4.2} & 84.4\textsubscript{3.6} & 88.3\textsubscript{3.5}
& 83.7\textsubscript{2.7} & 83.2\textsubscript{4.0} & 84.9\textsubscript{2.6} & 85.7\textsubscript{2.2} & 84.7\textsubscript{3.9} & 90.9\textsubscript{3.5} & 94.4\textsubscript{2.9} \\

12 
& 76.2\textsubscript{2.3} & 75.0\textsubscript{3.5} & 75.9\textsubscript{2.3} & 78.4\textsubscript{1.8} & 77.6\textsubscript{3.5} & 86.9\textsubscript{2.0} & 90.2\textsubscript{2.2}
& \textbf{84.3\textsubscript{1.9}} & \textbf{83.8\textsubscript{2.9}} & \textbf{85.4\textsubscript{2.0}} & \textbf{86.3\textsubscript{1.8}} & \textbf{85.3\textsubscript{2.8}} & \textbf{92.0\textsubscript{2.9}} & \textbf{95.3\textsubscript{2.5}} \\

16 
& 77.6\textsubscript{2.4} & 76.7\textsubscript{3.1} & 77.1\textsubscript{2.7} & 79.8\textsubscript{1.9} & 79.1\textsubscript{3.1} & 88.0\textsubscript{2.6} & 91.1\textsubscript{2.3}
& 85.4\textsubscript{2.8} & 85.1\textsubscript{4.1} & 86.6\textsubscript{2.6} & 87.1\textsubscript{2.2} & 86.4\textsubscript{4.0} & 93.3\textsubscript{3.0} & 95.8\textsubscript{2.9} \\

\bottomrule
\end{tabular}
}
\end{table*}

\subsection{\rqone}

\subsubsection{Approach}

To evaluate \textsc{FlaXifyer}'s classification performance, we conduct experiments varying two factors: (1) the pre-trained embedding model and (2) the number of shots per category.

\textbf{Encoders.} We compare two encoders with near parameter parity (109M vs. 125M), enabling controlled comparison:
\begin{itemize}
    \item \textbf{BGE}: \texttt{BAAI/bge-base-en-v1.5} (109M parameters), a state-of-the-art general-purpose text encoder trained on large-scale web data using contrastive learning~\cite{xiao_c-pack_2024}.
    \item \textbf{CodeBERT}: \texttt{microsoft/codebert-base} (125M parameters), a bimodal encoder pre-trained on programming languages and natural language~\cite{feng_codebert_2020}. CodeBERT has shown promise for flaky test classification tasks~\cite{akli_flakycat_2023}, making it a natural candidate for logs containing test/code snippets.
\end{itemize}

\textbf{Shot settings.} We evaluate $N \in \{2, 4, 8, 10, 12, 16\}$ shots per category, yielding training sets ranging from 26 to 208 examples with $K=13$ categories. The values 2, 4, 8, and 16 follow standard few-shot learning conventions~\cite{tunstall_efficient_2022, akli_flakycat_2023}, while 10 and 12 provide finer granularity in the intermediate range.

For each encoder-shot combination, we train and evaluate models using the MCCV procedure described in Section~\ref{sec:experimental_setup}, resulting in $2~\text{encoders} \times 6~\text{shot settings} \times 5~\text{hpp search trials} \times 30~\text{iterations} = 1800$ trained models. We report Macro F1 as the primary metric, along with MCC, and Top-k accuracy scores.

\subsubsection{Results}

Table~\ref{tab:setting_comparison} presents model performance across both encoders and the six shot settings.

\textbf{\textsc{FlaXifyer} achieves 84–85\% Macro F1 at 12–16 shots with BGE, which consistently outperforms CodeBERT across all shot settings, suggesting that general-purpose encoders are better suited for job logs.} At 12 shots, BGE reaches 84.3\% Macro F1 compared to 76.2\% for CodeBERT: a difference of 8.1 percentage points. This gap persists across all metrics: MCC (83.8\% vs. 75.0\%), precision (85.4\% vs. 75.9\%), recall (86.3\% vs. 78.4\%), and Top-2 accuracy (92.0\% vs. 86.9\%). Unlike flaky test classification, where code-aware models have shown promise~\cite{akli_flakycat_2023}, intermittent job failure logs prove to be better represented with general-purpose language models. This result underlines that job logs are not dominated by code or documentation fragments. Instead, they contain a heterogeneous mix of infrastructure messages, error traces, and tool-specific outputs that general-purpose encoders capture more effectively.

\textbf{With BGE, performance plateaus around 10–12 shots, with 12 shots achieving the best trade-off between performance, stability, and labeling effort.} Using the BGE encoder, F1-score increases from 66.1\% at 2 shots to 82.2\% at 8 shots, then stabilizes: 83.7\% at 10 shots, 84.3\% at 12 shots, and 85.4\% at 16 shots. The improvement from 12 to 16 shots (1.1 percentage points) is not statistically significant based on the Wilcoxon signed-rank test ($p > 0.05$). Moreover, variance decreases as shots increase up to 12, with standard deviation dropping from 6.2\% at 2 shots to 1.9\% at 12 shots. Overall, this suggests that 12 shots represent a practical labeling budget for deploying \textsc{FlaXifyer} with high and stable performance.

\textbf{Top-2 accuracy reaches 92.0\% at 12 shots, indicating strong practical utility even when the top prediction is incorrect.} With BGE at 12 shots, the correct category appears among the top two predictions in 92.0\% of cases. This metric is particularly relevant for diagnosis workflows: when developers are presented with two candidate categories, they can quickly verify against the log content and route the failure to the appropriate team or directly proceed with its resolution when possible.

\subsection{\rqtwo}

\subsubsection{Approach}

Our prior study~\cite{aidasso_diagnosis_2025} grouped the 13 categories by RFM priority rank. Ranks 1--8 constitute top/high priority categories (representing the most recent, frequent, and costly failures) which require immediate attention, while ranks 9--13 are medium priority categories representing natural candidates for incremental consideration. Hence, we simulate extending a deployed classifier by training models with increasing numbers $K$ of categories:

\begin{itemize}
    \item $K=8$: Top and high priority categories only (ranks 1--8)
    \item $K=10$: Core categories + 2 medium-priority (ranks 1--10)
    \item $K=13$: All 13 selected categories
\end{itemize}
This design reflects the practical scenario where an organization deploys a classifier for the most critical failures first, then progressively extends coverage to additional categories.

Based on RQ\textsubscript{1} results, we select the best-performing encoder and shot setting (BGE, 12 shots). For each category subset, we train models independently from the base pre-trained encoder using the MCCV procedure described in Section~\ref{sec:experimental_setup}. 

 We report Macro F1 to assess overall performance trends as $K$ increases. To examine whether adding medium-priority categories degrades performance on the core categories, we compare per-class F1-score ($\overline{F1}_k$) for core categories (ranks 1--8) across all three configurations of $K$. We also report per-class F1-score for newly added categories to assess how quickly they reach stable performance.

\subsubsection{Results}

\begin{table}[t]
\caption{Per-class F1-score across category set sizes $K$.}
\label{tab:f1_ensemble_k}
\centering
\begin{tabular}{lccc}
\toprule
\textbf{Category} & \textbf{$K{=}8$} & \textbf{$K{=}10$} & \textbf{$K{=}13$} \\
\midrule
misconfigured\_env\_variable        & 93.5\textsubscript{3.4}  & 89.6\textsubscript{8.4}  & 85.4\textsubscript{7.3}  \\
job\_execution\_timeout             & 85.0\textsubscript{9.3}  & 86.7\textsubscript{8.3}  & 84.6\textsubscript{8.9}  \\
dependency\_installation\_failure   & 92.6\textsubscript{6.1}  & 91.6\textsubscript{9.0}  & 92.4\textsubscript{2.9}  \\
runner\_pod\_waiting\_timeout       & 98.6\textsubscript{1.9}  & 98.2\textsubscript{2.0}  & 98.8\textsubscript{1.6}  \\
api\_gateway\_deployment\_error     & 92.0\textsubscript{3.9}  & 79.3\textsubscript{10.0} & 77.8\textsubscript{8.9}  \\
container\_registry\_server\_error  & 83.5\textsubscript{7.0}  & 83.4\textsubscript{6.6}  & 82.5\textsubscript{5.5}  \\
git\_transient\_error               & 96.1\textsubscript{2.8}  & 87.8\textsubscript{3.1}  & 87.9\textsubscript{2.7}  \\
flaky\_ui\_test                     & 98.6\textsubscript{1.1}  & 97.6\textsubscript{1.4}  & 97.3\textsubscript{1.1}  \\
external\_file\_invalid\_format     & --                       & 73.5\textsubscript{13.8} & 74.1\textsubscript{11.7} \\
host\_resolution\_failure           & --                       & 68.2\textsubscript{6.1}  & 68.4\textsubscript{5.8}  \\
runner\_image\_pull\_failure        & --                       & --                       & 95.4\textsubscript{3.4}  \\
remote\_call\_timeout               & --                       & --                       & 73.5\textsubscript{7.8}  \\
helm\_resource\_error               & --                       & --                       & 78.1\textsubscript{19.8} \\
\midrule
\textbf{Macro F1}                   & \textbf{92.5}            & \textbf{85.6}            & \textbf{84.3}            \\
\bottomrule
\end{tabular}
\end{table}

Table~\ref{tab:f1_ensemble_k} compares per-class F1-scores as the classifier is extended from $K=8$ core to $K=13$ total categories.

\textbf{\textsc{FlaXifyer} scales gracefully, with Macro F1 decreasing from 92.5\% to 84.3\% when extending from 8 to 13 categories.} This 8.2 percentage point decrease is expected as adding categories increases classification complexity. Notably, the transition from $K{=}10$ to $K{=}13$ incurs minimal degradation (85.6\% $\rightarrow$ 84.3\%), suggesting that the primary complexity increase occurs when first introducing medium-priority categories. To understand this degradation, we examine per-class performance across configurations.

\textbf{Most core categories remain stable as new categories are introduced, though two show notable degradation.} Five of the eight core categories maintain F1-scores within 2 percentage points across all $K$: \texttt{runner pod waiting timeout} (98.6\% $\rightarrow$ 98.8\%), \texttt{flaky ui test} (98.6\% $\rightarrow$ 97.3\%), \texttt{dependency installation failure} (92.6\% $\rightarrow$ 92.4\%), \texttt{job execution timeout} (85.0\% $\rightarrow$ 84.6\%), and \texttt{container registry server error} (83.5\% $\rightarrow$ 82.5\%). However, \texttt{api gateway deployment error} drops from 92.0\% to 77.8\% ($-$14.2), and \texttt{git transient error} decreases from 96.1\% to 87.9\% ($-$8.2). Both categories share semantic overlap with newly added network-related categories (\texttt{host resolution failure}, \texttt{remote call timeout}), making discrimination more challenging.

\textbf{At $K{=}13$, per-class F1 ranges from 68.4\% to 98.8\%, revealing distinct easy and difficult categories.} The easiest categories exhibit distinctive log patterns: \texttt{runner pod waiting timeout} (98.8\%), \texttt{flaky ui test} (97.3\%), \texttt{runner image pull failure} (95.4\%), and \texttt{dependency installation failure} (92.4\%), all with low variance (1.1--3.4\%). The most difficult categories are precisely those causing the degradation observed above: \texttt{host resolution failure} (68.4\%), \texttt{remote call timeout} (73.5\%), and \texttt{external file invalid format} (74.1\%). These categories share ambiguous network-related log patterns with others, as discussed in Section~\ref{sec:discussion}. Notably, \texttt{helm resource error} achieves moderate F1 (78.1\%) but exhibits the highest variance (19.8\%), highlighting inconsistent failure patterns within this category.

\begin{table}[t]
\centering
\caption{Relevance scoring for \textsc{LogSift} evaluation.}
\label{tab:relevance_scoring}
\begin{tabular}{lcccl}
\toprule
\textbf{Score} & \textbf{C1} & \textbf{C2} & \textbf{C3} & \textbf{Interpretation} \\
\midrule
High & $\checkmark$ & $\checkmark$ & $\checkmark$ & Explains prediction and actionable \\
Relevant & $\checkmark$ & $\checkmark$ & $\times$ & Explains prediction \\
Partial & $\checkmark$ & $\times$ & -- & Failure signal only \\
Not Rel. & $\times$ & -- & -- & No failure information \\
\bottomrule
\end{tabular}
\end{table}

\subsection{\rqthree}

\subsubsection{Approach}

We apply \textsc{LogSift} to a full test set split (614 log instances) using the best-performing model (BGE, 12 shots). We set the minimum segment size $\tau=2$ to ensure sufficient context for interpretation. Unlike classification, \textsc{LogSift}'s utility does not depend on prediction correctness. In fact, developers will use highlighted log segments to verify predictions, whether correct or not.

Therefore, we assess \textsc{LogSift} outputs along four dimensions: (1) \textbf{Reduction Ratio}, the average proportion of log statements eliminated while preserving the prediction; (2) \textbf{\textit{n}-Consistency}, the proportion of job instances where \textsc{LogSift} consistently reduces the log to $n$ lines or fewer, measuring practical reviewability; (3) \textbf{Execution Time}, how quickly \textsc{LogSift} highlights relevant segments, informing efficiency compared to manual verification; and (4) \textbf{Qualitative Relevance}, whether identified segments contain meaningful failure indicators.

For qualitative analysis, we randomly sampled 100 jobs and assessed each corresponding \textsc{LogSift} output against three binary criteria: \textbf{C1} (failure indicator), contains error messages, exceptions, or stack traces; \textbf{C2} (prediction-explanatory), explains the prediction, whether by supporting a correct classification or revealing why a misclassification occurred; and \textbf{C3} (actionable), contains specific repair information such as URLs or variable names. Each output is assigned a relevance score following Table~\ref{tab:relevance_scoring}.

\begin{table}[t]
\centering
\caption{\textsc{LogSift} quantitative evaluation results.}
\label{tab:logsift_results}
\begin{tabular}{lr}
\toprule
\textbf{Metric} & \textbf{Value} \\
\midrule
Mean reduction ratio & 74.4\% $\pm$ 27.6\% \\
30-consistency & 63.1\% \\
10-consistency & 40.9\% \\
2-consistency & 12.0\% \\
Mean execution time & 0.98s \\
\bottomrule
\end{tabular}
\end{table}

\begin{figure}[t]
\centering
\includegraphics[width=\linewidth]{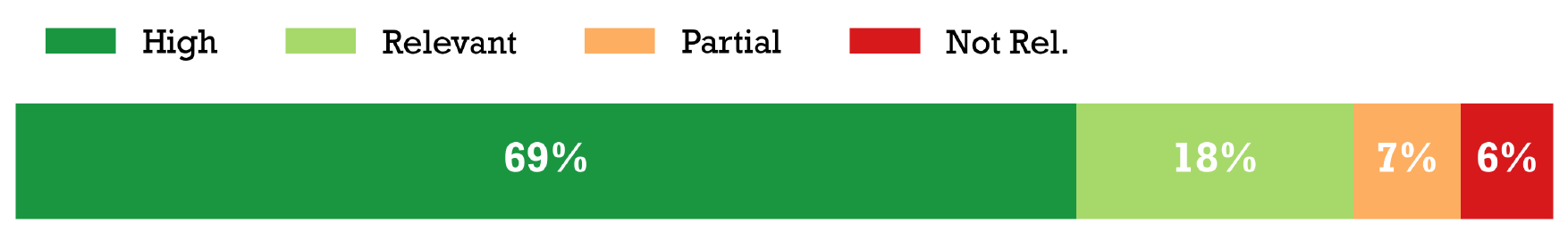}
\caption{\textsc{LogSift} relevance assessment results.}
\label{fig:logsift_relevance}
\end{figure}

\subsubsection{Results}Table~\ref{tab:logsift_results} summarizes the quantitative evaluation results, while Fig.~\ref{fig:logsift_relevance} presents the relevance assessment. Figures~\ref{fig:logsift_example},~\ref{fig:logsift_misclassified}, and~\ref{fig:logsift_not_relevant} illustrate representative \textsc{LogSift} outputs, with each header showing the predicted failure category and reduction ratio.

\textbf{\textsc{LogSift} effectively isolates influential statements, reducing log review effort by 74.4\% on average while maintaining sub-second execution time.} With a mean reduction ratio of 74.4\%, \textsc{LogSift} reduces the average 405-line log to approximately 104 lines. In the best cases, such as the output example in Fig.~\ref{fig:logsift_example} of a 359-line log, developers need only review 2 log statements (99.4\% reduction). More importantly, 30-consistency reaches 63.1\%, meaning that nearly two-thirds of logs are reduced to 30 lines or fewer, representing a practical threshold for quick manual review. For 40.9\% of log instances, outputs are 10 lines or fewer (a quick glance), and 12.0\% are narrowed down to pinpoint 2-line outputs. Mean execution time of 0.98 seconds per log ensures \textsc{LogSift} can be integrated into CI/CD pipelines without introducing any significant latency.

\begin{figure}[t]
\centering
\begin{logsegment}[\texttt{misconfigured\_env\_variable} \hfill 359 $\rightarrow$ 2 lines]
Building dependency release=broker-app, chart=app
in ./app.config.yaml: failed to render values files "app.values.yaml": failed to render [app.values.yaml], because of template: stringTemplate:5:15:...: required env var `IMAGE_NAME` is not set
\end{logsegment}
\caption{\textsc{LogSift} reduces a 359-line log to 2 lines, directly identifying the missing environment variable (\texttt{IMAGE\_NAME}).}
\label{fig:logsift_example}
\end{figure}

\begin{figure}[t]
\centering
\begin{logsegment}[\texttt{api\_gateway\_deployment\_error} \hfill 155 $\rightarrow$ 8 lines]
upload_zipfile_to_cloud_portal
test791.zip
curl: (26) Failed to open/read local data from file/application
zip file validate API is failed
200
Publish message to PubSub success
{"messageIds":["1403319032815554"]}200
section_end:1705665980:step_script
\end{logsegment}
\caption{\textsc{LogSift} output for a misclassified job (true: \texttt{external\_file\_invalid\_format}). Despite the incorrect prediction, the highlighted segment reveals the actual cause: a zip file read failure (\texttt{curl: (26)}) during upload, enabling developers to override the prediction.}
\label{fig:logsift_misclassified}
\end{figure}

\textbf{Qualitative analysis confirms that \textsc{LogSift} identifies meaningful failure indicators in 87\% of cases.} As shown in Fig.~\ref{fig:logsift_relevance}, 69\% of \textsc{LogSift} outputs received a \textit{High} relevance score, while an additional 18\% were rated \textit{Relevant}. Figure~\ref{fig:logsift_example} illustrates \textsc{LogSift} at its best: for a 359-line log of a \texttt{misconfigured env variable} failure, \textsc{LogSift} successfully retrieved the 2 lines pinpointing the missing variable (\texttt{IMAGE\_NAME}) and its context, precisely the information needed for repair. \textsc{LogSift} also provides relevant information for misclassifications: Figure~\ref{fig:logsift_misclassified} shows a job incorrectly predicted as \texttt{api gateway deployment error}, when the {true category} is \texttt{external file invalid format}. Despite the misclassification, \textsc{LogSift} surfaced the \texttt{curl} failure reading a zip file, enabling developers to still identify the underlying issue and override the prediction. However, \textsc{LogSift} does not always succeed, yielding 6\% outputs rated as \textit{Not relevant}. Figure~\ref{fig:logsift_not_relevant} shows a \texttt{host resolution failure} job where \textsc{LogSift} identified cache-related statements unrelated to DNS resolution. We hypothesize that such cases (6\% of evaluated instances) occur when failure signals are distributed throughout the log or when the model relies on spurious correlations.

\begin{figure}
\centering
\begin{logsegment}[\texttt{host\_resolution\_failure} \hfill 2,296 $\rightarrow$ 2 lines]
Creating cache app/resource/cachename...
.cache/pip: found 513 matching files and directories
\end{logsegment}
\caption{Example of \textsc{LogSift} producing a non-relevant segment. Despite a high reduction, the identified statements contain no failure indicators or host resolution information.}
\label{fig:logsift_not_relevant}
\end{figure}

\section{Discussion}
\label{sec:discussion}

\textbf{\textsc{FlaXifyer} and \textsc{LogSift} offer a practical, low-effort solution for automated failure triage and diagnosis.} With BGE at 12 shots, \textsc{FlaXifyer} achieves 84.3\% Macro F1 and 92.0\% Top-2 accuracy, meaning developers can confidently act on predictions or quickly verify between two candidates. The few-shot approach requires only 12 labeled examples per category, minimizing manual effort when extending to new failure types. If a new category becomes a priority, practitioners can collect 12 representative examples and retrain without extensive data collection. \textsc{FlaXifyer} predictions execute in milliseconds, enabling real-time feedback and automated triage. \textsc{LogSift} complements classification by reducing cognitive load. With a mean execution time under one second, it can run as a post-execution step for failed jobs, surfacing relevant log segments that accompany triage notifications or are saved as job artifacts for review. Together, these tools enable a streamlined workflow where failures are automatically categorized, directed to the appropriate team, and accompanied by actionable diagnostic information.

\textbf{Semantic overlap between failure categories explains both the value of Top-2 accuracy and \textsc{LogSift}'s limitations in certain cases.} Several categories exhibit overlapping log patterns: environment variable errors during container operations, file format issues during API gateway deployment, DNS resolution failures during Git operations, and test timeouts during UI rendering. This overlap arises partly from label consolidation in our prior work~\cite{aidasso_diagnosis_2025}, where related failure patterns were grouped to avoid label explosion. For instance, file read failures were consolidated under \texttt{external file invalid format}, and various Helm-related errors under \texttt{helm resource error}, resulting in heterogeneous log patterns within these categories. Such overlap explains why Top-2 accuracy (92.0\%) substantially exceeds Top-1 accuracy (85.3\%), as the correct category often appears among the top two predictions. However, this same overlap causes \textsc{LogSift} to occasionally surface segments that match the predicted rather than the true category, as observed in 6\% of evaluated instances. Future work could explore multi-label classification, where logs are mapped to multiple applicable categories, though Top-2 prediction already mitigates this limitation in practice. Additionally, some failures require diagnosis beyond job logs. For example, a file read failure may stem from intermittent infrastructure issues (e.g., network filesystem timeouts) not captured in application logs. Future studies should investigate the use of complementary data sources, such as infrastructure logs or telemetry data, to support end-to-end root cause analysis.

\section{Threats to Validity}
\label{sec:threats}

\textbf{Internal Validity.} Our dataset relies on automated labeling using regex patterns, achieving 91\% precision~\cite{aidasso_diagnosis_2025}. While this introduces approximately 9\% label noise, few-shot learning has demonstrated robustness to noisy labels~\cite{tunstall_efficient_2022}. Few-shot performance can also vary depending on which examples are sampled for training. We mitigate this using Monte Carlo Cross-Validation with 30 iterations to ensure stable estimates. Hyperparameter selection using Optuna~\cite{akiba_optuna_2019} with 5 trials may not identify globally optimal configurations, though our results demonstrate strong performance across settings. For qualitative evaluation, the first author assessed \textsc{LogSift} outputs using structured binary criteria, with findings validated during weekly meetings with engineers at TELUS. Sample judgments are included in our replication package \cite{aidasso_artifact_2026} for verification.

\textbf{External Validity.} Our evaluation focuses on industrial projects at TELUS, which may limit generalizability to other organizations. However, the failure categories studied (e.g., network timeouts, dependency failures, infrastructure errors) are common across CI/CD systems~\cite{lampel_when_2021, durieux_empirical_2020}, and the few-shot approach requires only 12 labeled examples per category, facilitating adaptation to new contexts. To support reproducibility, our replication package \cite{aidasso_artifact_2026} includes the complete \textsc{FlaXifyer} and \textsc{LogSift} implementation, experimental scripts for all RQs, and sanitized evaluation results. The TELUS dataset is not included due to confidentiality constraints. For replication in new environments, we include our regex-based labeling tool~\cite{aidasso_flakeranker_2025} along with a sample dataset of labeled intermittent job failures from the \texttt{veloren/veloren} GitLab project.

\section{Conclusion and Future Work}
\label{sec:conclusion}

Intermittent job failures disrupt CI/CD pipelines, wasting computational resources and requiring significant diagnosis effort \cite{aidasso_diagnosis_2025, aidasso_efficient_2025, olewicki_towards_2022, lampel_when_2021}. In this paper, we introduced \textsc{FlaXifyer}, a few-shot learning approach for predicting intermittent job failure categories using pre-trained language models. Evaluated on industrial projects at TELUS, \textsc{FlaXifyer} achieves 84.3\% Macro F1 and 92.0\% Top-2 accuracy with only 156 labeled log examples for 13 categories (12 per category), and scales gracefully when extending to new failure categories. We also proposed \textsc{LogSift}, an interpretability technique that reduces log review effort by 74.4\% while surfacing relevant failure information in 87\% of cases. Together, these tools enable automated triage, reduce cognitive load, and accelerate failure resolution.

As part of our future work, we intend to explore multi-label classification to better handle failures exhibiting overlapping log patterns, along with more refined category definitions to reduce semantic ambiguity. We also plan to investigate AI agents for automated failure resolution, leveraging \textsc{FlaXifyer}'s predictions and \textsc{LogSift}'s highlighted segments to trigger category-specific repair actions without human intervention.  This direction contributes to our broader vision of holistic and continuous build optimization~\cite{aidasso_towards_2025}.

\begin{acks}
We acknowledge the support of the Natural Sciences and Engineering Research Council of Canada (NSERC), ALLRP/ 576653-2022. This work was also supported by TELUS and Mitacs through the Mitacs Accelerate program. 
\end{acks}

\balance
\bibliographystyle{ACM-Reference-Format}
\bibliography{references}

\appendix

\end{document}